\documentclass[floats,floatfix,showpacs,amssymb,prd,twocolumn,superscriptaddress,nofootinbib,nolongbibliography,reprint]{revtex4-2}

\usepackage{amssymb,amsmath,verbatim,mathtools,needspace,enumitem,etoolbox,graphicx,physics,microtype,afterpage,bigints,gensymb,tabularx,xspace}

\usepackage[dvipsnames, usenames]{xcolor}
\definecolor{linkcolor}{rgb}{0.0,0.3,0.5}
\definecolor{dodgerblue}{HTML}{1E90FF}
\usepackage[unicode, colorlinks=true, linkcolor=linkcolor, citecolor=linkcolor, filecolor=linkcolor,urlcolor=linkcolor, pdfusetitle]{hyperref}
\usepackage[all]{hypcap}
\usepackage[T1]{fontenc}
\usepackage[utf8]{inputenc}
\usepackage{orcidlink}

\makeatletter
\newcommand*{\balancecolsandclearpage}{\close@column@grid \cleardoublepage \twocolumngrid}
\makeatother

\newcommand{\milan}{\affiliation{Dipartimento di Fisica ``G. Occhialini'', Universit\'a degli Studi di Milano-Bicocca, Piazza della Scienza 3, 20126 Milano, Italy}}
\newcommand{\infn}{\affiliation{INFN, Sezione di Milano-Bicocca, Piazza della Scienza 3, 20126 Milano, Italy}}
\newcommand{\caltech}{\affiliation{TAPIR 350-17, California Institute of Technology, 1200 E California Boulevard, Pasadena, CA 91125, USA}}

\begin{document}

\title{Population inference of eccentric binary black holes\\using asymptotic constants of motion}

\author{Chiara Anselmo$\,$\orcidlink{0009-0008-9328-1177}}
\email{c.anselmo@campus.unimib.it}

\milan \infn

\author{Giulia Fumagalli$\,$\orcidlink{0009-0004-2044-989X}}

\caltech

\author{Davide Gerosa$\,$\orcidlink{0000-0002-0933-3579}}

\milan \infn


\date{\today}

\begin{abstract}
In standard gravitational-wave population analyses of compact binaries, source parameters are defined at a reference frequency. For quantities that evolve during the inspiral, this means constructing a population of systems at different evolutionary stages, which complicates the astrophysical interpretation of the results.
The past-time infinity, or zero-frequency, limit provides a mathematically well-defined reference where all sources enter the population fit on equal footing. This approach was first proposed for spin tilts, which evolve due to spin precession but converge to constant values as binaries are back-propagated.
The orbital eccentricity also evolves throughout the inspiral and approaches unity in the zero-frequency limit for all binaries, making it unsuitable as a common reference point for population analyses. On the other hand, the Newtonian orbital angular momentum is an asymptotic constant of motion of the relativistic two-body problem and remains finite in this limit.
In this paper, we make use of this known asymptotic behavior to reformulate eccentric population inference in terms of orbital angular momentum rather than eccentricity and demonstrate this approach on simulated gravitational-wave data.
Our work constitutes the first example of eccentric population inference using only asymptotic constants of motion. 
Importantly, after the inference has been performed using this consistent reference, the results can be propagated to any preferred frequency for comparison with astrophysical predictions.
\end{abstract}

\maketitle

\section{Introduction}

As the catalog of gravitational-wave (GW) detections by the LIGO/Virgo/KAGRA (LVK) interferometers continues to grow~\cite{2019PhRvX...9c1040A, 2021PhRvX..11b1053A, 2023PhRvX..13d1039A, 2025arXiv250818082T, 2026arXiv260527225T}, more information can be gained on the properties of the astrophysical population that produces these events. The standard approach is hierarchical Bayesian inference~\cite{2019MNRAS.486.1086M, 2022hgwa.bookE..45V}: individual events are first analyzed separately to obtain posterior distributions for their source parameters, and these posteriors are subsequently combined to infer the parameters describing the underlying population, commonly referred to as hyperparameters.

The implicit assumption in such hierarchical analyses is that the source parameters entering the population model should represent the same physical quantity for every event. However, in the case of compact binaries in GW astronomy, sources evolve while being observed. Source parameters are typically defined at a detector-frame reference frequency, such as $f_{\rm ref} = 20~\rm Hz$ in LVK observations~\cite{2026arXiv260527226T}.
We argue that, when designing a population fit, one should make use of parameters that are constants of motion of the relativistic two-body problem. This is the case, for instance, for the objects' masses. By contrast, for quantities that evolve during the inspiral, defining parameters at a fixed detector-frame frequency implies that one is constructing a population of binaries sampled at different stages of their evolution, which significantly complicates any downstream (astro)physical interpretation.

For example, this is true for black hole (BH) spin orientations, which precess as the binary evolves toward merger~\cite{1994PhRvD..49.6274A}. While the second post-Newtonian (2PN) order spin-precession equations admit a non-trivial constant of motion, namely the effective spin~\cite{2008PhRvD..78d4021R}, the tilt angles between the spins and the binary angular momentum evolve on a timescale shorter than that of the inspiral.
Spin orientations are believed to be highly informative observables for discriminating between BH binary formation channels~\cite{2021hgwa.bookE..16M,2022PhR...955....1M} and, as such, are commonly included in GW population analyses~\cite{2026arXiv260527226T}.
While the spin angles are not constants of motion in general, they are asymptotic constants of motion in the Newtonian limit~\cite{2015PhRvD..92f4016G,2022PhRvD.106b3001J,2023PhRvD.108b4042G}, i.e., they tend to constant values as $t \to -\infty$, or equivalently $f_{\rm ref} \to 0~\rm Hz$.

Exploiting this property, \citeauthor{2022PhRvD.105b4076M}~\cite{2022PhRvD.105b4076M} proposed using the zero-frequency limit as a common reference point for population analyses. The key idea is to evolve each single-event posterior distribution from $f_{\rm ref}$ to 0~Hz, impose a population distribution there and fit the data, and then forward-evolve the inferred population from 0~Hz to $f_{\rm ref}$, where the binaries are observable.
This approach is now widely used in GW astronomy, with flagship LVK data releases providing spin-tilt information at 0~Hz~\cite{ligo_scientific_collaboration_and_virgo_2026_17602505}. Of course, astrophysical binaries do not literally exist at past infinity, but instead form at some finite time. Nevertheless, 0~Hz can be used as a formal reference point in parameter space at which the tilt angles represent the same physical quantity for all binaries. Results from such population fits can then be presented at a different time or frequency, depending on the targeted application.

Orbital eccentricity is arguably an even more promising observable for distinguishing among formation channels. Eccentricity decays as the binary inspirals~\cite{1964PhRv..136.1224P}, which is a blessing for astrophysical inference: an eccentricity measurement that confidently departs from zero unambiguously indicates that the binary must have formed recently through some form of dynamical interaction. At the same time, this circularization property of relativistic binaries is a curse for hierarchical Bayesian fits: orbital eccentricity does not admit a useful zero-frequency limit, and one cannot use the same argument as for spin orientations. Under orbit-averaged GW-driven evolution, binaries become wider and more eccentric when propagated backward in time, such that $e \rightarrow 1$ as $f_{\rm ref}\rightarrow 0~\rm Hz$~\cite{1964PhRv..136.1224P}. In practice, non-adiabatic effects become important at large eccentricities~\cite{2021PhRvD.103f4066T,2025PhRvD.112b4012F}. Population analyses with eccentricity are under active development~\cite{2024PhRvD.110f3009Z,2026PhRvD.113l3025Z,2026arXiv260329019G, 2026arXiv260322461M, 2026arXiv260715453Z}, but all current efforts include eccentricity information at some nominal, detector-frame reference frequency, thus suffering from the caveats discussed above.

While eccentricity approaches the same asymptotic value for all binaries, the Newtonian orbital angular momentum is an asymptotic constant of motion that retains information about the binary's orbital properties.
In the zero-frequency limit, the semi-major axis $a$ diverges and the eccentricity $e$ tends to unity, but their combined evolution leaves the orbital angular momentum $L \propto \sqrt{a(1-e^2)}$ finite~\cite{1964PhRv..136.1224P,2023PhRvD.108l4055F}. Although this asymptotic behavior has long been understood from the classical evolution of eccentric binaries, its implications for hierarchical population inference have not been explored. This observation motivates reformulating eccentric population inference in terms of orbital angular momentum rather than eccentricity. We adopt the proposed approach by constructing a population model for the asymptotic orbital angular momentum and demonstrate its feasibility with simulated data.

\section{Formulation}
\label{sec:formulation}

\subsection{Orbital angular momentum at past infinity}

In Newtonian gravity, the orbital angular momentum of a binary on an elliptical orbit is
\begin{equation}
\label{newtonang}
\frac{L}{M^2} = \frac{q}{(1+q)^2}\sqrt{\frac{a}{M}(1-e^2)} \ ,
\end{equation}
where $M = m_1 + m_2$ is the total mass, $q = m_2/m_1 \leq 1$ the mass ratio, $a$ the semi-major axis, and $e$ the eccentricity. Hereafter, we use geometrical units with $c = G = 1$. In the absence of dissipation effects, the orbital angular momentum is conserved.

Gravitational radiation causes both the orbital separation and the eccentricity to evolve with time. As the binary inspirals, the orbit shrinks and progressively circularizes. The leading-order, orbit-averaged evolution of $a$ and $e$ is described by the well-known equations first derived by \citeauthor{1964PhRv..136.1224P}~\cite{1964PhRv..136.1224P}:
\begin{align}
\label{eq:Peters1}
\frac{\dd a}{\dd t}&=-\frac{64}{5}\,\frac{q\, M^3}{(1+q)^2}\,\frac{1+\frac{73}{24}e^2+\frac{37}{96}e^4}{a^3 \left(1-e^2\right)^{7/2}} \ ,\\
\label{eq:Peters2}
\frac{\dd e}{\dd t}&=-\frac{304}{15}\,\frac{q \,M^3}{(1+q)^2}\frac{e \left(1+\frac{121}{304}e^2\right)}{a^4 \left(1-e^2\right)^{5/2}} \ .
\end{align}
An important caveat is that the orbit-averaged approximation underlying Eqs.~(\ref{eq:Peters1}--\ref{eq:Peters2}) becomes unreliable for highly eccentric orbits, particularly at small semi-major axes~\cite{2021PhRvD.103f4066T,2025PhRvD.112b4012F}.
We make use of the equations above because the objective here is to investigate the asymptotic behavior of the system rather than accurately describe its dynamics. The approach to GW populations outlined in this paper can be generalized to include non-adiabatic effects in the source propagation.

We use Kepler's third law $M = (2\pi f_{\rm orb})^2 a^3$ and further assume that the dominant gravitational emission frequency is twice the orbital frequency $f_{\rm orb}$, such that the detector-frame reference frequency is $f_{\rm ref} = 2f_{\rm orb}/(1+z)$.
This is an approximation, as binaries with $e \gtrsim 0.3$ emit more power at higher harmonics of the orbital frequency~\cite{2003ApJ...598..419W,2021RNAAS...5..275H,2024ApJ...969..132V}. Although straightforward, such a generalization is beyond the scope of this proof-of-concept study.
Under this approximation, we express the orbital semi-major axis $a$ as a function of the GW frequency $f_{\rm ref}$:
\begin{equation} 
\label{afref}
\frac{a}{M} = \big[\pi f_{\rm ref} M (1+z)\big]^{-2/3} \ ,
\end{equation}
which is the expression used throughout the paper when propagating binaries across frequencies.

From Eqs.~(\ref{eq:Peters1}-\ref{eq:Peters2}), it is straightforward to show that, as $t \to -\infty$ (or, equivalently, $f_{\rm ref}\to 0$~Hz), one has $a \to \infty$ and $e \to 1$. However, these limits are approached with $a \propto (1-e^2)^{-1}$, such that the Newtonian angular momentum of Eq.~(\ref{newtonang}) remains finite. Integrating  Eqs.~(\ref{eq:Peters1}-\ref{eq:Peters2}) from the initial condition $(a_0,e_0)$ yields the orbital angular momentum along the inspiral,
\begin{align}
\label{L_a0e0}
\frac{L}{M^2} &= \frac{q}{(1 + q)^2}~\sqrt{\frac{a_0}{M}} \notag \\ 
&\times \sqrt{\frac{ (1 - e_0^2)~e^{12/19}(304 + 121e^2)^{870/2299}}
{e_0^{12/19}(304 + 121e_0^2)^{870/2299}}} \ ,
\end{align}
here parametrized as a function of $e$. 
The values of $L$ at 0~Hz, hereafter denoted as $L_{\infty}$, can be obtained by taking the limit $e \rightarrow 1$ in Eq.~(\ref{L_a0e0}):
\begin{equation}
\label{L0Hz_a0e0}
\frac{L_{\infty}}{M^2} = \frac{q}{(1 + q)^2}~\sqrt{\frac{a_0}{M} \frac{ (1 - e_0^2)~425^{870/2299}}
{e_0^{12/19}(304 + 121e_0^2)^{870/2299}}
} \ .
\end{equation}
Thus, although both the semi-major axis and eccentricity become degenerate in the zero-frequency limit, each reference configuration specified by $(a_0,e_0)$ remains associated with a unique, finite value of $L_\infty$, and we can define our population model at 0~Hz directly in terms of $L_\infty$.

\begin{figure*}
\centering
\includegraphics[width=\textwidth]{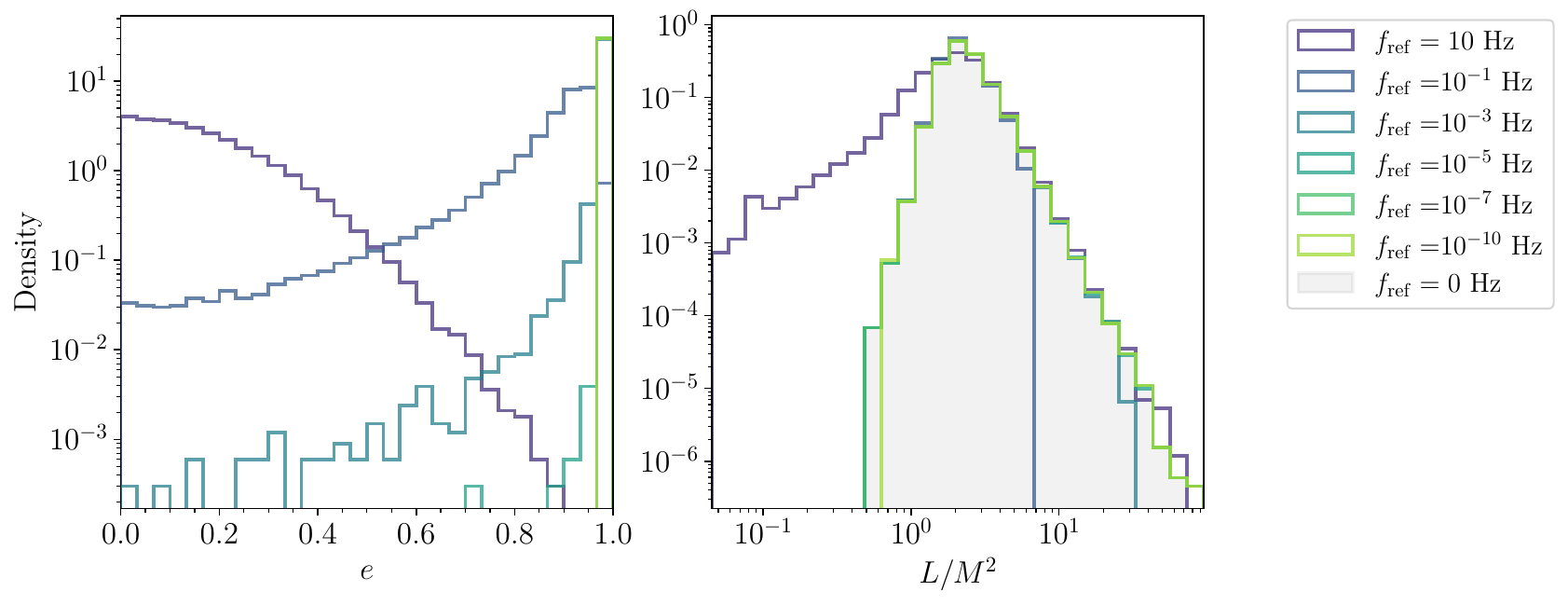}
\caption{Assuming a truncated Gaussian distribution in eccentricity at $f_{\rm ref}=10$~Hz, we show the resulting distributions as samples are propagated to progressively lower frequencies, expressed in terms of both eccentricity (left panel) and orbital angular momentum (right panel). While the eccentricity distribution becomes singular and infinitely concentrated at $e=1$, the orbital angular momentum distribution converges to a regular probability density function. The latter overlaps with the distribution of $L_\infty$ computed using Eq.~\eqref{L0Hz_a0e0} (shaded region).
For illustrative purposes, in this figure we fixed $M=50~M_\odot$, $q=1$, and $z=0.1$.
}
\label{fig:pop_prop}
\end{figure*}

The left panel of Fig.~\ref{fig:pop_prop} shows the result of propagating an initial eccentricity distribution from 10~Hz to progressively lower frequencies. For this example, we fix $M = 50~M_{\odot}$, $q = 1$, and $z = 0.1$, and assume a truncated Gaussian distribution $\mathcal{N}(e | \mu = 0,\sigma = 0.2)$ for the eccentricity at 10~Hz, with $e \in [0,1]$.
As the frequency approaches 0~Hz, the eccentricities pile up near 1, making a population model defined at such frequencies ill-posed. In the right panel of Fig.~\ref{fig:pop_prop}, the same eccentricity values are mapped to values of the orbital angular momentum. As binaries are back-propagated to past infinity, the distributions of $L$ converge to a well-behaved limiting distribution, which is given by $L_\infty$ from Eq.~\eqref{L0Hz_a0e0}.

\subsection{Population model}
\label{popmodels}

Our toy population consists of non-spinning binary BHs, characterized by their source-frame total mass $M$, mass ratio $q$, redshift $z$, and asymptotic orbital angular momentum $L_{\infty}$. While this provides a suitable four-dimensional problem to illustrate the key point of this paper, the framework presented here can be extended to include additional binary parameters, such as spins and tidal deformabilities.

Referring to $\theta = \{M,q,z,L_\infty\}$ as the single-event parameters and to $\lambda = \{\alpha, M_{\rm min}, M_{\rm max}, \beta, \kappa, s, \gamma \}$ as the population parameters, we neglect correlations among source parameters for simplicity and assume that the full population model $p_{\rm pop}(\theta \mid \lambda)$ factorizes into four components:
\begin{enumerate}
\item
For the total mass, we adopt a truncated power-law distribution, 
\begin{equation}
p_{\rm pop}(M \mid \alpha, M_{\rm min}, M_{\rm max}) \propto M^\alpha \,, \;\;  M_{\rm min} \leq M \leq M_{\rm max} \ .
\end{equation}
\item The mass ratio is also modeled with a power-law distribution,
\begin{equation}
p_{\rm pop}(q \mid \beta) \propto q^\beta\,. 
\end{equation} 
\item
The redshift distribution is taken to be
\begin{equation}
p_{\rm pop}(z \mid \kappa)
\propto
\frac{dV_c}{dz}
(1+z)^{\kappa -1} \ , \quad z \leq 1.5 \ ,
\end{equation}
where $V_c$ is the comoving volume and the upper redshift limit is chosen to lie beyond the detector's horizon. The expression above corresponds to a merger-rate density proportional to $(1+z)^{\kappa}$~\cite{2015ApJ...806..263D,2018ApJ...863L..41F}.
\item We model $L_{\infty}$ with a lognormal distribution,
\begin{equation}
\label{ppopLinf}
p_{\rm pop}(L_{\infty} \mid s, \gamma)
= \frac{1}{s L_{\infty} \sqrt{2\pi}}
\exp\left[
-\frac{\ln^2(L_{\infty}/\gamma)}{2s^2}
\right]\,,
\end{equation}
where $s>0$ and $\gamma$ are the shape and scale parameters, respectively, and $L_{\infty} > 0$.
The reasoning behind this modeling choice is discussed in Appendix~\ref{app:ppopL}.
\end{enumerate}

\subsection{Simulated sources}
\label{sec:sim_sources}

We inject sources from a population with true hyperparameters (loosely inspired by the values reported in Ref.~\cite{2026arXiv260527226T})
$\alpha = 2$,
$M_{\rm min}= 10~M_{\odot}$,
$M_{\rm max}= 250~M_{\odot}$,
$\beta =1$,
$\kappa =2.5$,
$s= 0.5$,
and $\gamma = 2.4$.
In a realistic analysis, these parameters would be unknown and inferred from the data.
From the corresponding population distribution, we sample a set of sources, each described by the source parameters $\theta$. Notably, single-event parameter estimation (PE) using current waveform models does not provide $L_\infty$ but rather some PN notion of eccentricity at a reference frequency~\cite{2023PhRvD.108l4063R,2025PhRvD.112l1503A,2026arXiv260901568G}.
Irrespective of the specific notion of eccentricity adopted by each waveform approximant, the key point is that PE samples are expressed in terms of a different parameter set, $\theta' = \{ M,q,z,e \}$, where, for the toy model developed here, $e$ denotes the eccentricity at $f_{\rm ref}=10~\rm Hz$. When generating injections from a population expressed in terms of $L_{\infty}$, we convert the samples to $e$ using the inverse mapping described in Appendix~\ref{app:mapping}. This conversion would not be necessary when analyzing real data.

We select the sources that are observed by the detector by applying the detection probability $p({\rm det} \mid \theta')$, defined in the next subsection. This subset of the simulated population represents the observed GW catalog. For this study, we consider $N_{\rm obs} = 150$ events.

To simulate measurement uncertainties, we perturb the true value of each parameter of the observed sources by drawing
\begin{equation}
\theta'_{\rm obs} \sim \mathcal{N}(\theta'_{\rm true},\sigma_{\theta'}) \ .
\end{equation}
We adopt a phenomenological measurement uncertainty that scales inversely with the signal-to-noise ratio (SNR)~\cite{1995PhRvD..52..848P},
\begin{equation}
\label{theta'}
\sigma_{\theta'} = \frac{C_{\theta'}}{{\rm SNR}}\ ,
\end{equation}
with
$C_M = 50~M_{\odot}$, $C_q = 1$, $C_z = 2$, and $C_e = 1$. These coefficients are chosen to yield measurement uncertainties of the order expected for moderately loud signals. In particular, our choice for eccentricity gives $\sigma_e \simeq 0.05$ for a source with ${\rm SNR}\simeq20$, comparable in order of magnitude to the eccentricity sensitivity reported in previous LVK analyses~\cite{2018PhRvD..98h3028L}.

For each observed signal, we generate a synthetic PE posterior as a Gaussian distribution centered at $\theta'_{\rm obs}$ with standard deviation $\sigma_{\theta'}$. The single-event posteriors are truncated at the physical boundaries of the source parameters.

\subsection{Selection effects}
\label{sec:sel_effs}
An important aspect of population inference is accounting for selection effects~\cite{2019MNRAS.486.1086M,2022hgwa.bookE..45V}, which introduce biases between the intrinsic and observed distributions of sources.
In a realistic analysis, the detection probability would be obtained from injection campaigns using the same search pipelines employed to detect eccentric binary BHs. As with the PE samples, this probability would be provided in terms of the in-band source parameters $\theta'$, yielding $p({\rm det} \mid \theta')$ rather than $p({\rm det} \mid \theta)$. It would depend jointly on all source parameters, accounting for correlations between eccentricity and the other binary properties. 

Given the lack of a suitable injection campaign covering eccentricity and the pedagogical purpose of this study, we adopt a simplified model in which selection effects are factorized into a component depending only on $(M,q,z)$, computed assuming quasi-circular sources, and a universal eccentricity-dependent correction:
\begin{equation}
p({\rm det} \mid M,q,z,e) = p({\rm det} \mid M,q,z,e=0) \times p({\rm det} \mid e) \ .
\end{equation}
This approach neglects, among other effects, the known degeneracy between chirp mass and eccentricity~\cite{2021PhRvD.104l4021B,2022PhRvD.105b3003F}. As the eccentricity increases, the signal morphology transitions from a chirp to a sequence of repeated bursts, requiring correspondingly different detection strategies that are not captured by this toy model.

\begin{figure}
\centering
\includegraphics[width=0.8\columnwidth]{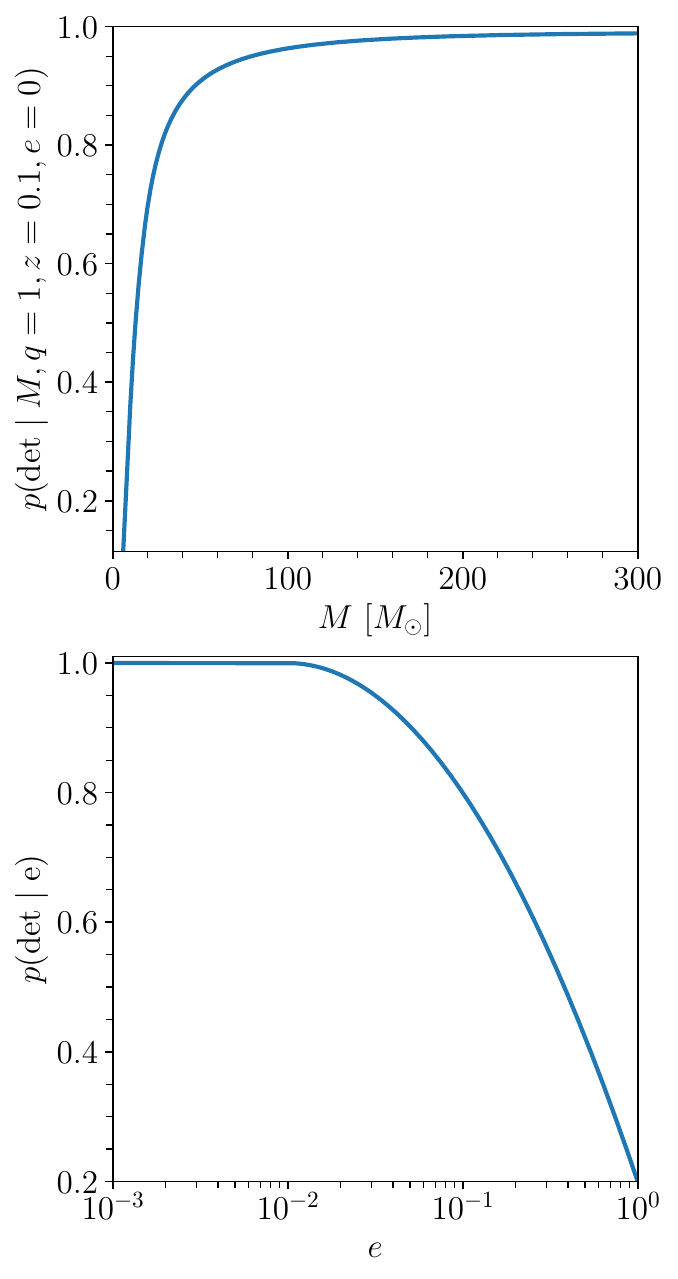}
\caption{Approximation of the selection function for eccentric, non-spinning binary BHs. The top panel shows the quasi-circular component $p({\rm det} \mid M,q=1,z=0.1,e=0)$ as a function of the total mass $M$. The bottom panel shows the eccentric correction $p({\rm det} \mid e)$ as a function of the eccentricity $e$ measured at a reference frequency of $10$ Hz.
}
\label{fig:pdets}
\end{figure}

For the circular component, the detection probability is computed by thresholding the optimal SNR, as implemented in the Python module \textsc{gwdet}~\cite{gwdet}. 
We consider a single detector with a LIGO ``design'' sensitivity curve~\cite{LIGOcurve}, and use the IMRPhenomD waveform model~\cite{2016PhRvD..93d4007K} with an SNR threshold of 8. We analytically average over sky location, inclination, and polarization angles~\cite{1993PhRvD..47.2198F,2024CQGra..41l5002G}.
The top panel of Fig.~\ref{fig:pdets} shows $p({\rm det} \mid M, q=1, z=0.1, e=0)$ as a function of $M$.

The eccentricity-dependent contribution $p({\rm det} \mid e)$ is approximated from the recovered fraction of eccentric sources shown in the inset of Fig.~1 of Ref.~\cite{2021ApJ...921L..43Z}.
We write it as a simple quadratic function of $\log_{10}e$:
\begin{equation}
p({\rm det} \mid e) = -0.2 \log^2_{10}e - 0.8 \log_{10}e + 0.2 \ ,
\end{equation}
which is shown in the bottom panel of Fig.~\ref{fig:pdets}.
This approximation is broadly consistent with the semianalytic sensitivity estimates obtained in Ref.~\cite{2026arXiv260329019G}, which account for correlations with other parameters (bottom panel of their Fig.~4), and those obtained in Ref.~\cite{2026arXiv260614045V}, which are restricted to non-spinning, low-eccentricity binary neutron star systems (bottom panel of their Fig.~6). Specifically, the correction tends to 1 as $e \to 0$, so that the overall detection probability recovers the quasi-circular result, and decreases with increasing eccentricity, reflecting the difficulty of detecting burst-like signals from eccentric binaries.

\subsection{Hierarchical inference}
\label{sec:hierarchical}

Assuming an inhomogeneous Poisson process and marginalizing over the event rate, the population posterior distribution is given by~\cite{2019MNRAS.486.1086M,2022hgwa.bookE..45V}
\begin{equation}
\label{hier}
p(\lambda \mid d) \propto \pi(\lambda) \prod_{i=1}^{N_{\rm obs}} \frac{ \displaystyle \int p_{\rm pop}(\theta \mid \lambda)  \frac{ p(\theta' \mid d_i) }{ \pi(\theta')} \left| \frac{\partial\theta}{\partial\theta'}\right|  \dd \theta'}{ \displaystyle \int p_{\rm pop}(\theta \mid \lambda)  p(\det \mid \theta')\left| \frac{\partial\theta}{\partial\theta'}\right|   \dd \theta'} \ ,
\end{equation}
where $\pi(\lambda)$ is the hyperprior, $p(\theta' \mid d_i)$ is the single-event posterior distribution, and $\pi(\theta')$ is the PE prior used to obtain that posterior.

For simplicity, we assume that all PE analyses were performed using uniform priors in $\theta'$.
Since the population model $p_{\rm pop}(\theta\mid \lambda)$ is formulated in terms of the source parameters $\theta$ (which include $L_\infty$), one must account for the change of variables from $\theta$ to $\theta'$ (which include $e$ instead) through the corresponding Jacobian. In our simplified formulation, the absolute value of the Jacobian determinant $\left|\partial \theta/\partial \theta'\right|$ reduces to $\left|\partial L_\infty/\partial e\right|$, which can be computed analytically (see Appendix~\ref{app:mapping}). In a more realistic case that includes BH spins, evaluating this Jacobian would likely require numerical integration of the PN equations of motion~\cite{2021arXiv210610291K,2023PhRvD.108l4055F,2025PhRvD.112d4067P} or suitable fits.

The integrals in the numerator of Eq.~\eqref{hier} are estimated using Monte Carlo sums, with $1000$ samples drawn from each posterior $p(\theta' \mid d_i)$ generated as described in Sec.~\ref{sec:sim_sources}. The selection-effect integral in the denominator is evaluated by importance sampling over a fixed set of $5 \times 10^5$ analytically generated injections drawn from a uniform distribution that covers the entire hyperparameter space.

We sample $p(\lambda \mid d)$ using uniform priors with
$\alpha\in [1,3]$,
$M_{\rm min} \in[6,30]~M_{\odot}$,
$M_{\rm max} \in [200,300]~M_{\odot}$,
$\beta \in [0,2]$,
$\kappa \in [1,6]$,
$s \in [10^{-1},2]$,
and
$\gamma \in [1,3]$
.
We adopt the \textsc{emcee}~\cite{2013PASP..125..306F} sampler with 100 walkers, each running for $10^4$ steps. We evaluate the autocorrelation lengths $\tau$ and their maximum $\tau_{\rm max}$, discard the first $2 \times \tau_{\rm max}$ steps as burn-in, and thin the chain by a factor of $\tau_{\rm max}/2$. The resulting chain contains $\sim 10^4$ hyperparameter samples.

\section{Results}
\label{sec:results}

\subsection{Reconstructed population in $\boldsymbol{L_\infty}$}

\begin{figure*}[hbtp]
\centering
\includegraphics[width=\textwidth]{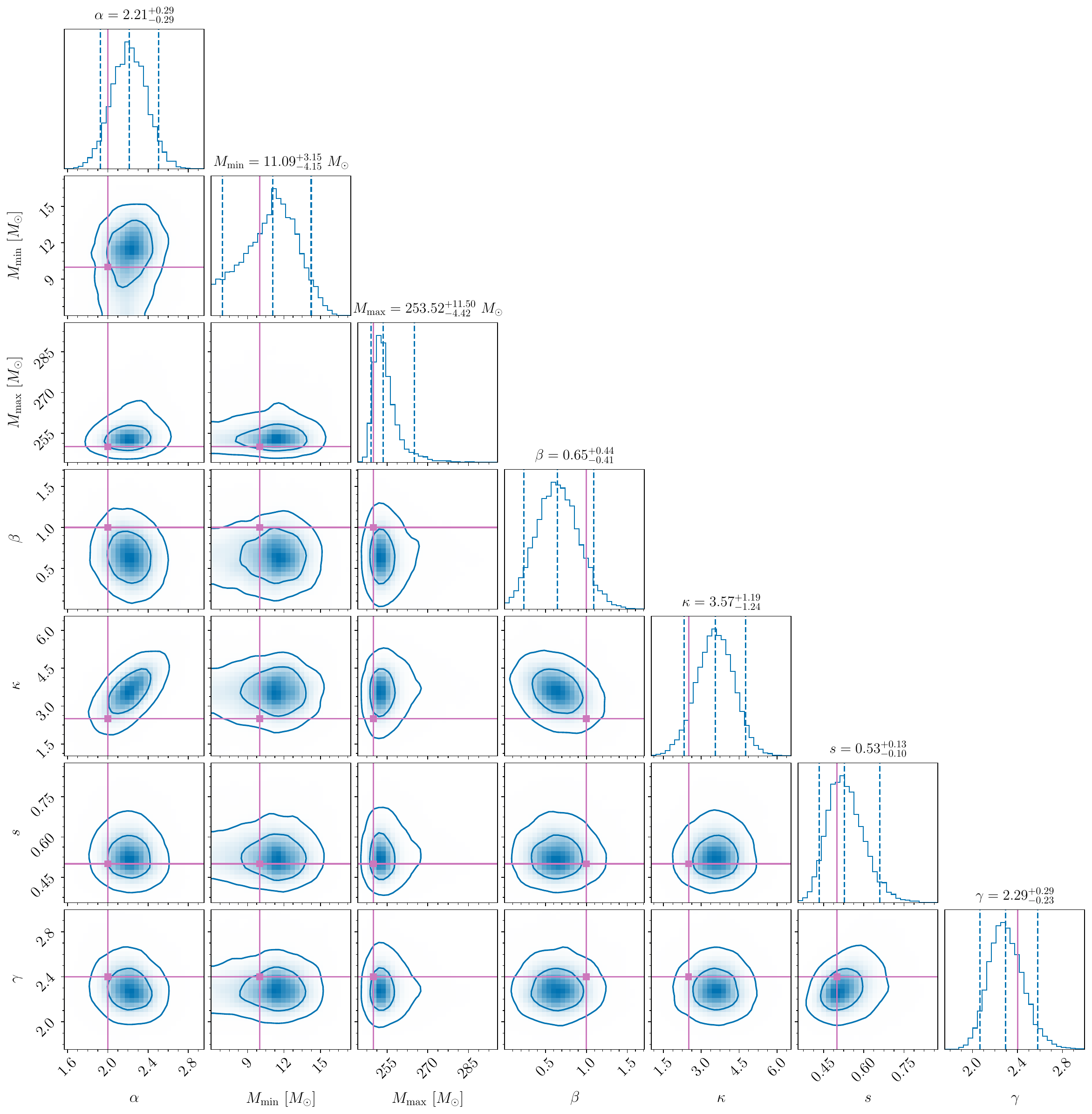}
\caption{Posterior distributions of the population parameters for a fit performed at past time infinity using the asymptotic angular momentum. Contours indicate the 50\%  and 90\% credible regions.
The pink lines correspond to the true values used to generate the simulated data.
}
\label{fig:corner}
\end{figure*}

We present the recovered joint distribution of the seven hyperparameters of our model in Fig.~\ref{fig:corner}. For all population marginal distributions, the true value is contained within the 90\% credible interval.

\begin{figure*}[hbtp]
\centering
\includegraphics[width=0.8\textwidth]{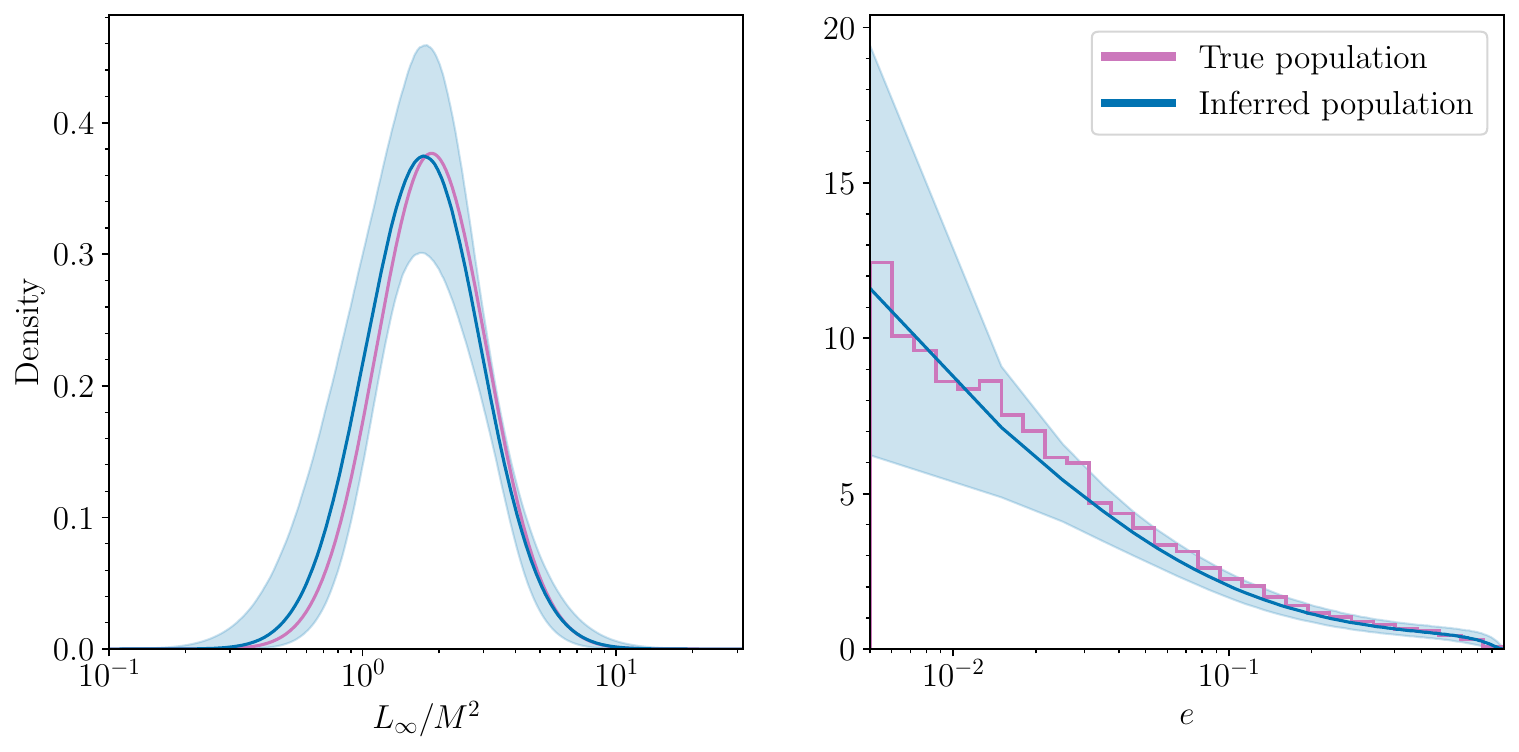}
\caption{Reconstructed distribution of the asymptotic angular momentum $L_\infty$ (left) and the eccentricity $e$ at $f_{\rm ref}= 10~\rm Hz$ (right). The solid blue curves show the median PPDs, and the shaded regions indicate the 90\% pointwise credible bands. The pink curve (left) and histogram (right) show the population model evaluated at the true hyperparameters.
}
\label{fig:ppds}
\end{figure*}

The posterior predictive distribution (PPD) for $L_\infty$ is shown in the left panel of Fig.~\ref{fig:ppds} against its true population model $p_{\rm pop}(L_\infty \mid s_{\rm true}, \gamma_{\rm true})$. To obtain this, we randomly draw $N=10^4$ samples of $(s,\gamma)$ from the hyperposterior chain and evaluate the corresponding $p_{\rm pop}(L_\infty \mid s,\gamma)$ on a fixed $L_\infty$ grid. We then compute, at each grid point, the median and the $5^{\rm th}$ and $95^{\rm th}$ percentiles across the $N$ posterior draws. The inferred population is consistent with the distribution used to generate the simulated sources.

Let us stress that, compared to a population fit performed at a non-zero reference frequency, this fit uses (asymptotic) constants of motion of the binary dynamics, thus addressing the interpretation challenges described in the introduction. If desired, once the fit has been performed, the results can still be reported in terms of the binary parameters when the sources enter the sensitivity band of our detectors. For each hyperparameter posterior draw, we
\begin{enumerate}
\item generate $N$ samples of $(M,q,z,L_\infty)$ from our population model;
\item transform each sample to eccentricity at ${f_{\rm ref}=10~\rm Hz}$ (see Appendix~\ref{app:mapping} for details);
\item compute the pointwise median and 5th and 95th percentiles of the resulting eccentricity distribution.
\end{enumerate}
The PPD for $e$ at $10$~Hz is  shown in the right panel of Fig.~\ref{fig:ppds}, together with samples drawn from the population model using the true hyperparameters. The true population is well-contained inside the 90\% band.

\subsection{Divergence at low reference frequencies}
\label{sec:hyper_drift}

To further illustrate the need to use asymptotic constants of motion in population inference, we investigate how the inferred population parameters depend on the reference frequency when one insists on using eccentricity instead. To this end, we consider two synthetic populations, $P_1$ and $P_2$, each containing binaries whose asymptotic angular momentum $L_\infty$ follows a Gaussian distribution with means $\mu_1=2.0$, $\mu_2=4.0$ and standard deviations $\sigma_1=0.2$, $\sigma_2=0.3$. We consider a one-dimensional inference problem where, for each binary, we draw the source parameters from Gaussian distributions,
\begin{equation}
M \sim \mathcal{N}(40,15)~M_{\odot} \ , \ q \sim \mathcal{N}(0.7,0.15) \ , \ z \sim \mathcal{N}(0.3,0.15) \ ,
\end{equation}
truncated at their physical boundaries, and keep them fixed throughout the analysis. We then convert $L_\infty$ to the corresponding eccentricity at 10~Hz and generate the PE posteriors as described above.

We perform hierarchical population inference at a set of logarithmically spaced reference frequencies spanning $f_{\rm ref}\in[10^{-30},10]~\rm{Hz}$. At each reference frequency, we model the eccentricity population using a skewnormal distribution, parametrized by the shape, location, and scale parameters $(\rm{shape},\rm{loc},\rm{scale})$, following the convention of \texttt{scipy.stats.skewnorm}. 
We infer these three hyperparameters independently for each $f_{\rm ref}$.

\begin{figure*}[hbtp]
\centering
\includegraphics[width=\textwidth]{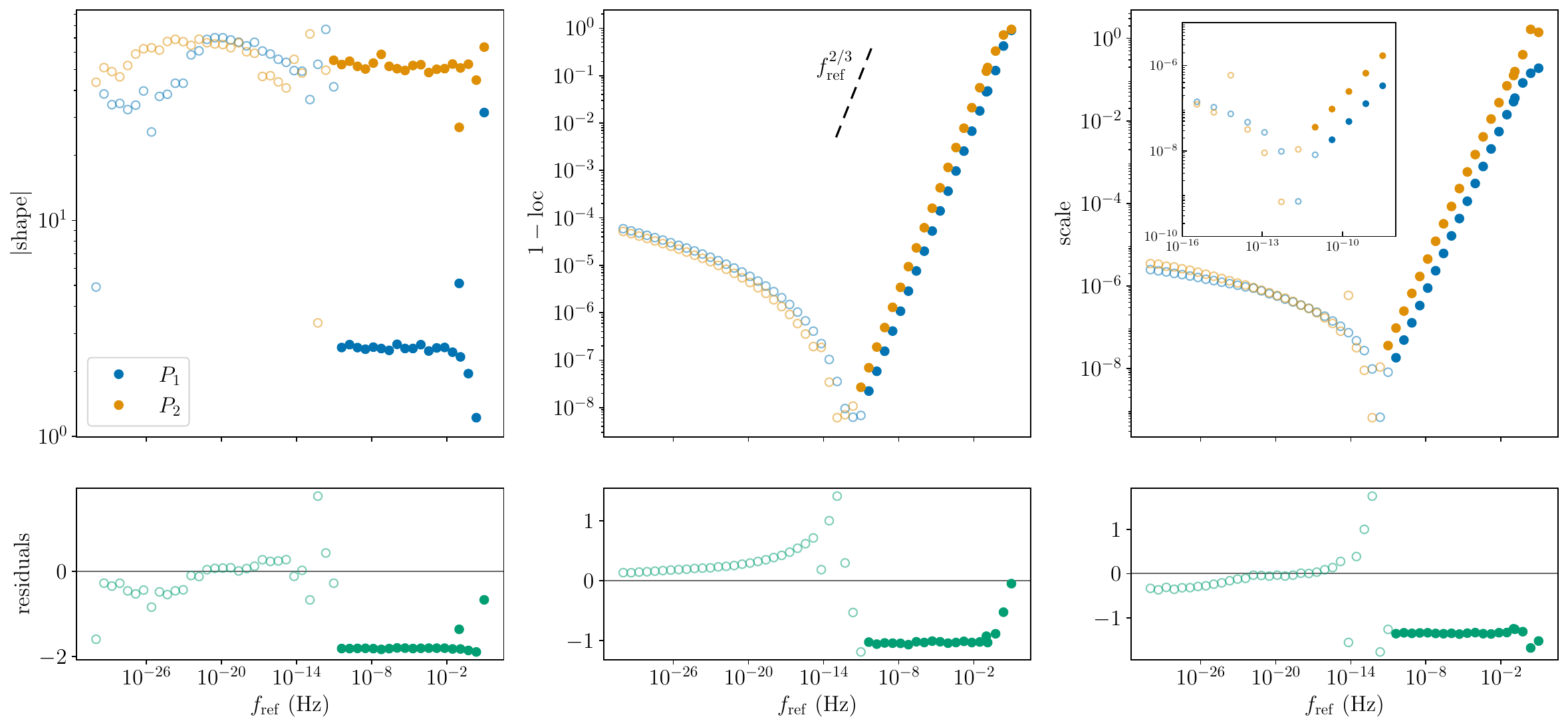} 
\caption{Dependence of the inferred hyperparameters on the reference frequency. The top panels show, from left to right, the reconstructed median values of the shape, location, and scale parameters of a skewnormal distribution in eccentricity. Blue and orange markers refer to the two simulated populations. Filled and open markers distinguish frequencies for which the fraction of samples outside the interpolation grid is below and above 1\%, respectively. The dashed line in the central panel indicates the expected $f_{\rm ref}^{2/3}$ scaling in the high-eccentricity limit.
The bottom panels show the relative difference between the median estimates for the two universes, defined as $2(P_1-P_2)/(P_1+P_2)$.
}
\label{fig:hyper_drift}
\end{figure*}

To obtain the eccentricity at $f_{\rm ref}$, we evolve the PE eccentricity samples using Peters' equations~(\ref{eq:Peters1}--\ref{eq:Peters2}). In particular, we solve
\begin{equation}
g[e(f_{\rm ref})] =
\frac{a(f_{\rm ref})}{a(10~\rm{Hz})}
g[e(10~\rm{Hz})] \ ,
\end{equation}
where
\begin{equation}
g(e)=
\frac{e^{12/19}}{1-e^2}
\left(1+\frac{121}{304}e^2\right)^{870/2299} \ .
\end{equation}
We evaluate the inverse mapping numerically by interpolating $\log g(e)$ on a grid of eccentricity values between 0 and 1, logarithmically spaced near the edges.
The evolution equations also provide an expectation for the low-frequency behavior. From Kepler's third law, $a(f_{\rm ref})\propto f_{\rm ref}^{-2/3}$~\eqref{afref}. In the high-eccentricity limit, the function $g(e)$ scales as $g(e)\propto (1-e)^{-1}$. Then, at sufficiently low frequencies,
\begin{equation}
1-e(f_{\rm ref})\propto f_{\rm ref}^{2/3}
\end{equation}
up to a binary-dependent factor.
Figure~\ref{fig:hyper_drift} shows the inferred hyperparameters $(\rm{shape},\rm{loc},\rm{scale})$ as a function of $f_{\rm ref}$. Specifically, as a criterion to mark numerical instabilities, we distinguish frequencies for which more than 1\% of the samples lie outside the interpolation grid. Residuals are computed as the relative difference between the median estimates for the two populations, i.e., ${2~(P_1-P_2)/(P_1+P_2)}$, where $P_1$ and $P_2$ denote the corresponding median hyperparameter estimates.

The two injected distributions are different but, using eccentricity, they can only be distinguished at sufficiently high reference frequencies. At low reference frequencies, on the other hand, the inferred population parameters become increasingly difficult to resolve. One would erroneously infer that the two populations are the same, with significant numerical errors. This is because, as $f_{\rm ref}$ decreases, the mapping between eccentricity and orbital evolution becomes increasingly ill-conditioned: all eccentricity values tend to 1, and eccentricity thus loses its ability to distinguish between different populations.

For this controlled experiment, the lowest frequency at which the parameters can be reliably inferred is set by the finite extent and resolution of our interpolation grid. While one could brute-force this exercise to lower frequencies, the key point we wish to stress is that using eccentricity itself inevitably leads to a loss of discriminatory power in the zero-frequency limit. The asymptotic angular momentum $L_\infty$, on the other hand, does not suffer from this problem by construction, as it is an asymptotic constant of motion of the underlying binary dynamics.

\section{Conclusions}
\label{sec:conclusions}

A peculiarity of population inference in GW astronomy is that the sources whose population we wish to model evolve while being observed. Defining their parameters at a given reference frequency or time complicates the interpretation of the results, as one constructs a population of sources at different stages of their evolution. 

In this work, we explore a framework for hierarchical population inference of eccentric binary BHs based on the asymptotic orbital angular momentum $L_\infty$, rather than the orbital eccentricity at detection $e$. While the eccentricity evolves throughout the inspiral and approaches unity in the zero-frequency limit, making it unsuitable for defining a common reference point for population analyses, in contrast the Newtonian orbital angular momentum remains finite in this limit and can therefore serve as a common reference quantity across the population.

It is important to stress that performing the population fit at past infinity, or $f_{\rm ref}=0$~Hz, should be understood as a statistical procedure for obtaining a consistent population fit without the ambiguities described above. Astrophysical binaries form at a finite time and thus do not exist at $f_{\rm ref}=0$~Hz. Once the fit has been performed, model predictions can be propagated to whichever reference frequency or time is suitable for comparison with astrophysical models. 

One apparent drawback of this approach is the need to formulate the population model in terms of $L_\infty$, whereas physical intuition and current astrophysical predictions are more readily expressed in terms of $e$. However, a decade of GW astronomy has shown that the choice of population model remains crucial even for familiar physical quantities such as BH masses. A substantial body of work is devoted to understanding how inferred population properties depend on the adopted model, and flagship analyses still explicitly compare results obtained with different models to assess robustness~\cite{2026arXiv260527226T}. Related efforts aim to diagnose possible model misspecification using, e.g., posterior predictive checks~\cite{2022PASA...39...25R,2026arXiv260406090M}. In this context, the field is moving toward increasingly flexible, non-parametric approaches that allow the shape of the population distribution and possibly the number of parameters to be inferred from the data~\cite{2022MNRAS.509.5454R,2025PhRvD.111f3043H,2025ApJ...994L..52T}. Such approaches can be applied to $L_\infty$ just as they can to $e$.

In parallel, predictions from astrophysical models could also be expressed in terms of $L_\infty$, allowing us to build intuition for the expected behavior of this variable. Following this line of reasoning, comparisons between models and data could then be performed directly in a parameter space consisting solely of asymptotic constants of motion.

While this paper presents a proof of concept, our approach is readily generalizable.
First, although we considered non-spinning binary BHs, the extension to spinning systems is conceptually straightforward because the spin-orbit angles themselves are asymptotic constants of motion for quasi-circular binaries~\cite{2015PhRvD..92f4016G,2022PhRvD.106b3001J,2023PhRvD.108b4042G}, and different parametrizations can be constructed for eccentric sources~\cite{2023PhRvD.108l4055F}. Essentially, this would amount to combining the approach of Ref.~\cite{2022PhRvD.105b4076M} with that of this paper. Computational challenges might nevertheless arise from the cost of propagating spinning, eccentric binaries, although fast precession-averaged approaches are now available~\cite{2023PhRvD.108l4055F}.
Second, the treatment of the selection function presented here is approximate and should be improved. To this end, GW searches and PE pipelines for eccentric sources are being deployed increasingly widely.

For context, it is important to note that orbital eccentricity does not admit a unique definition in general relativity, and the definition used in waveform models typically differs from that used to report astrophysical predictions~\cite{2024ApJ...969..132V}. Already within the PN framework, the quasi-Keplerian parametrization introduces distinct radial, temporal, and angular eccentricities, which share the same Newtonian limit but differ through relativistic corrections~\cite{2019CQGra..36b5004L}. Definitions based on coordinate quantities can also inherit gauge dependencies, motivating the development of eccentricity estimators constructed directly from gravitational waveforms~\cite{2023PhRvD.108j4007S,2025PhRvD.111b4008B}.
Our approach sidesteps the need to select a universal finite-frequency eccentricity definition for population inference. Waveform models used for single-event PE and injection campaigns can retain their own eccentricity conventions, provided that the corresponding samples and probability densities are consistently mapped to the common asymptotic variable $L_\infty$.

To conclude, our results provide, for the first time, a strategy for performing eccentric population inference in GW astronomy using only asymptotic constants of motion, thus consistently placing all sources entering the population ``on the same page''. At the same time, this framework retains the ability to report predictions for the eccentricity distribution at observationally relevant frequencies.

\acknowledgements

We thank Katerina Chatziioannou, Eliot Finch, Matthew Mould, Isobel Romero-Shaw, Alexandre Toubiana, and the participants of the workshop ``{\it Binary compact objects on eccentric orbits and their gravitational-wave signatures}'' (Corfu, Greece, 2026) for discussions. 
C.A. and D.G. are supported by 
ERC Starting Grant No.~945155--GWmining, 
Cariplo Foundation Grant No.~2021-0555, 
Italian-French University (UIF/UFI) Grant No.~2025-C3-386,
MUR Grant ``Progetto Dipartimenti di Eccellenza 2023-2027'' (BiCoQ),
and the INFN TEONGRAV initiative.
G.F. is supported by the Sherman Fairchild Postdoctoral Fellowship at the California Institute of Technology.
D.G. is supported by MUR Young Researchers Grant No. SOE2024-0000125. 
Computational work was performed 
at CINECA with allocations through INFN and the University of Milano-Bicocca.

\appendix

\section{Population prior for $\boldsymbol{L_\infty}$}
\label{app:ppopL}

\begin{figure*}
\centering
\includegraphics[width=0.8\textwidth]{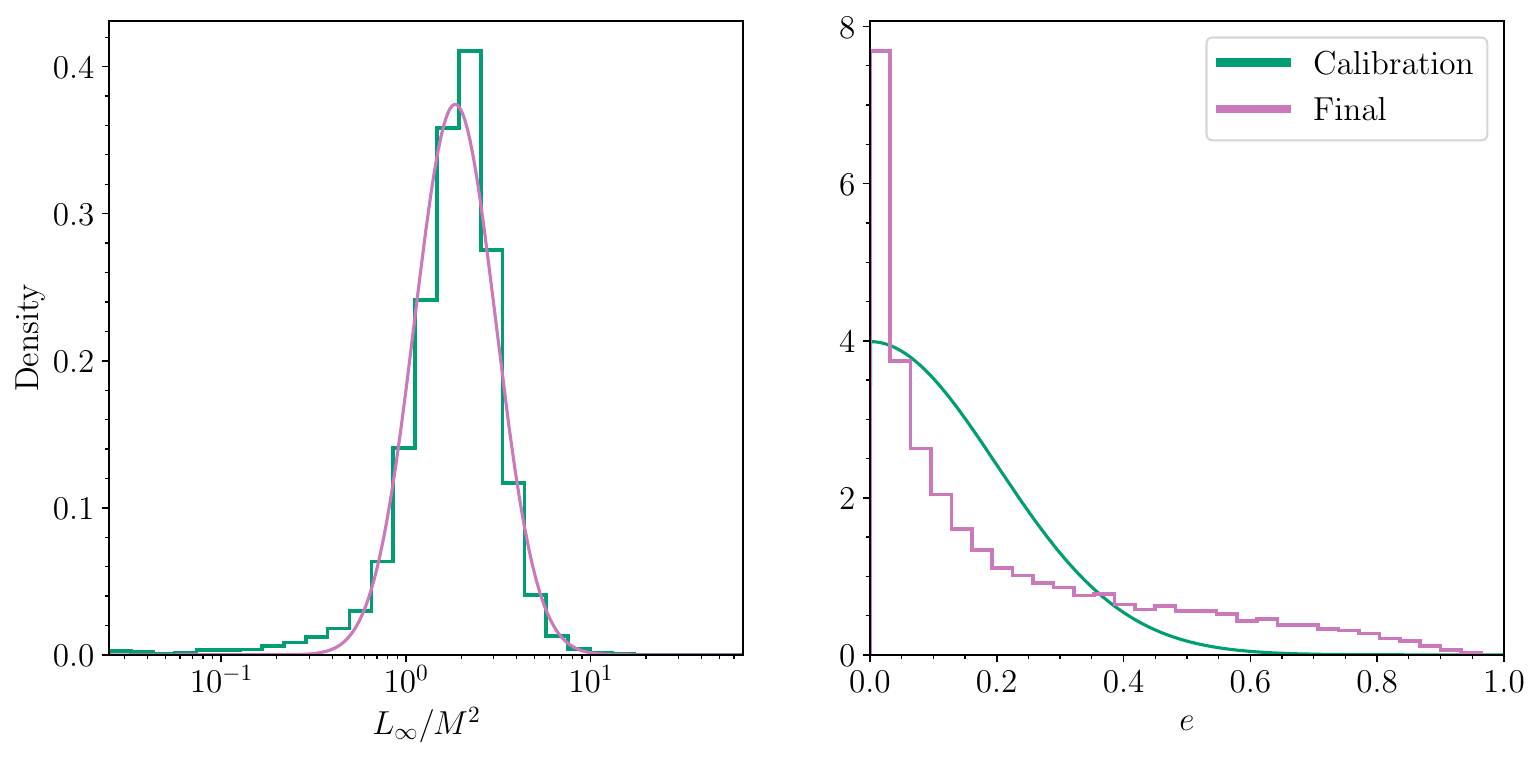} 
\caption{Distributions of asymptotic angular momentum $L_\infty$ (left) and eccentricity $e$ at $f_{\rm ref}=10$~Hz (right), obtained from the calibration population (green) and our fit (pink). Importantly, the calibration population is defined in terms of $e$ but fitted in terms of $L_\infty$, neglecting correlations with the other binary parameters.
}
\label{fig:ppopL_calib}
\end{figure*}

There are currently no established prescriptions for the population distribution of the asymptotic orbital angular momentum $L_\infty$, with astrophysical models most often presenting predictions in terms of the eccentricity $e$. The phenomenological distribution given in Eq.~\eqref{ppopLinf} was motivated by combining our population model for $M$, $q$, and $z$ (Sec.~\ref{popmodels}) with a truncated Gaussian distribution of $e$ at $f=10~\rm Hz$, centered at zero. In this paper, the latter serves as a putative astrophysical prediction expressed in terms of the eccentricity at detection. We then generated $5 \times 10^5$ sources characterized by $(M,q,z,e)$ and computed the corresponding values of $L_\infty$ using Eq.~\eqref{L0Hz_a0e0}. The resulting $L_\infty$ distribution is shown in the left panel of Fig.~\ref{fig:ppopL_calib} and appears to be well described by a lognormal distribution, which is the model we use in this paper.

An important consequence of this simplification is that the eccentricity values obtained when sampling the final population in $(M,q,z,L_\infty)$ do \emph{not} follow the original Gaussian distribution (see the right panel of Fig.~\ref{fig:ppopL_calib}). This is because the fitted distribution $p_{\rm pop}(L_{\infty} \mid s, \gamma)$ is defined independently of the other population parameters, even though it was calibrated using $L_\infty$ values computed from $(M,q,z,e)$. Therefore, transforming samples back to $e$ will not recover the initial eccentricity distribution. This would remain true regardless of the functional form chosen to fit the $L_\infty$ data.
Moreover, we emphasize that this calibration procedure is intended to provide a proof of concept, rather than to establish an astrophysically preferred population model for $L_\infty$.

\section{$\boldsymbol{e_0 \leftrightarrow L_{\infty}}$ mapping}
\label{app:mapping} 

Starting from Eq.~\eqref{L0Hz_a0e0}, we express $a_0$ in terms of $f_{\rm ref}$ using Eq.~(\ref{afref}) and isolate the square-root term that depends solely on $e_0$:
\begin{equation}
L_{\infty} \equiv A(M,q,z,f_{\rm ref}) \times B(e_0) \ .
\end{equation}
Defining
\begin{equation}
x \equiv \dfrac{L_{\infty}}{A(M,q,z,f_{\rm ref})} \ ,
\end{equation}
the inverse problem of determining $e_0$ given $L_{\infty}$ reduces to solving $B(e_0) = x$.
In our implementation, this equation is solved in logarithmic space, i.e., by inverting $\ln B(e_0)$.
The derivative required for the numerical inversion is
\begin{equation}
\frac{d\ln B}{de_0} = \frac{96 + 292e_0^2 + 37e_0^4}
{e_0\left(-304 + 183e_0^2 + 121e_0^4\right)} \ .
\label{dlogB}
\end{equation}
We then solve $\ln B(e_0) = \ln x$ using Newton's method:
\begin{equation}
e_0^{(n+1)} = e_0^{(n)} - \frac{ \ln B\left(e_0^{(n)}\right) - \ln\left[L_{\infty}/A(M,q,z,f_{\rm ref})\right]}{ \left. \dfrac{d\ln B}{de_0} \right|_{e_0=e_0^{(n)}} } \ .
\label{newton_e0}
\end{equation}

To initialize the iteration, we use the limiting behavior of $\ln B(e_0)$ near $e_0=0$ and $e_0=1$. For $e_0\rightarrow0$,
\begin{equation}
\ln B(e_0) \simeq - \frac{6}{19}\ln e_0 + \frac{435}{2299} \ln\left(\frac{425}{304}\right) \ ,
\end{equation}
while for $e_0\rightarrow1$,
\begin{equation}
\ln B(e_0) \simeq \frac{1}{2}\ln\left[2(1-e_0)\right] \ .
\vspace{0.5em}
\end{equation}
These asymptotic expressions, consistent with those derived in Ref.~\cite{1964PhRv..136.1224P}, provide the initial guess for the Newton iteration in the corresponding eccentricity regimes. The solution is kept within the physical interval $e_0 \in (0,1)$.

\bibliography{Lpop}

\end{document}